\documentstyle[amssymb,preprint,aps]{revtex}

\begin{document}
\draft
\title{{\bf General Approach to the Quantum Kicked Particle in a Magnetic 
Field: Quantum-Antiresonance Transition}}
\author{{\bf Itzhack Dana}$^{1}${\bf \ and Dmitry L. Dorofeev}$^{1,2}$}
\address{$^{1}$Minerva Center and Department of Physics, Bar-Ilan University,
Ramat-Gan 52900, Israel}
\address{$^{2}$Department of Physics, Voronezh State University, Voronezh 
394693, Russia}
\maketitle

\begin{abstract}
The quantum kicked particle in a magnetic field is studied in a weak-chaos
regime under realistic conditions, i.e., for {\em general} values of the
conserved coordinate $x_{{\rm c}}$ of the cyclotron orbit center. The system
exhibits spectral structures [``Hofstadter butterflies'' (HBs)] and quantum
diffusion depending sensitively on $x_{{\rm c}}$. Most significant changes
take place when $x_{{\rm c}}$ approaches the value at which quantum
antiresonance (exactly periodic recurrences) can occur: the HB essentially
``doubles'' and the quantum-diffusion coefficient $D(x_{{\rm c}})$ is
strongly reduced. An explanation of these phenomena, including an
approximate formula for $D(x_{{\rm c}})$\ in a class of wave packets, is
given on the basis of an effective Hamiltonian which is derived as a power
expansion in a small parameter. The global quantum diffusion of a
two-dimensional wave packet for all $x_{{\rm c}}$ is briefly considered.
\newline
\end{abstract}

\pacs{PACS numbers: 05.45.Mt, 05.45.Ac, 03.65.-w}

\begin{center}
{\bf I. INTRODUCTION}\\[0pt]
\end{center}

Simple quantum systems whose classical counterparts are nonintegrable
exhibit a variety of remarkable phenomena \cite{qc} which continue to
attract much interest in several contexts. A realistic system, whose
experimental realization by different methods has been the object of many
recent works (see a very partial list in Ref. \cite{erkr}), is the
well-known kicked rotor \cite{qc,kr,kp}, modelling the
dynamical-localization phenomenon. A class of systems exhibiting basically
different phenomena, such as quantum diffusion associated with a fractal
spectrum, are represented by Hamiltonians periodic in phase space \cite
{h,drh,dz,gkp,wa,ehb,cn,ls,gkp1,khm,d,d1,d2,dfw,drf,ff}. The simplest such
system is the integrable Harper model \cite{h} with Hamiltonian $H=\cos
(x)+\cos (p)$, whose quantization describes the energy spectrum within a
Bloch band in a magnetic field or within a broadened Landau level in a
crystalline periodic potential \cite{h,drh,dz,ehb}. Plotting this spectrum
as a function of the magnetic flux through a unit cell (this flux is related
to a scaled Planck constant for the Harper model) gives the well-known
``Hofstadter butterfly'' \cite{drh}, featuring a clearly fractal structure 
\cite{wa}. Quite recently \cite{ehb}, much of this structure was detected
experimentally.\newline

The simplest nonintegrable system with a mixed phase space and a periodic
Hamiltonian is the kicked Harper model (KHM) \cite
{cn,ls,gkp1,khm,d,d1,d2,dfw,drf} with $H=\cos (p)+\cos (x)\sum_{s=-\infty
}^{\infty }\delta (t/T-s)$, which reduces to the Harper Hamiltonian when the
time period $T\rightarrow 0$. The KHM models the regular-chaotic transition 
\cite{cn,drf} as well as the effect of chaos on a fractal spectrum and on
quantum diffusion \cite{gkp1}. It seems that the only realistic
interpretation of the KHM is its exact relation \cite{d1} with a particular
case of the two-dimensional (2D) system of the periodically kicked particle
in a magnetic field \cite{zas,da,pr}. This system is described by 
\begin{equation}
H=\Pi ^{2}/2-K\cos (x)\sum_{s=-\infty }^{\infty }\delta (t-sT),  \label{H}
\end{equation}
where ${\bf \Pi }=(\Pi _{x},\ \Pi _{y})={\bf p}-{\bf B\times r}/(2c)$ is the
kinetic momentum of a particle with unit charge and unit mass in a uniform
magnetic field ${\bf B}$ (along the $z$-axis) and $K$ is a nonintegrability
parameter. Let us first summarize some known facts about (\ref{H}) \cite
{zas,da,pr}. It is crucial to represent (\ref{H}) in the natural degrees of
freedom in a magnetic field, given by the independent conjugate pairs $(\Pi
_{x},\ \Pi _{y})$ and $(x_{{\rm c}},\ y_{{\rm c}})$ (coordinates of the
center of a cyclotron orbit) \cite{jl}. Defining $u=\Pi _{x}/\omega $,
$v=\Pi _{y}/\omega $, where $\omega =B/c$ is the cyclotron frequency, and
using the relation $x_{{\rm c}}=x+\Pi _{y}/\omega =x+v$ (from simple
geometry), (\ref{H}) can be rewritten as follows \cite{da} 
\begin{equation}
H=\omega ^{2}(u^{2}+v^{2})/2-K\cos (x_{{\rm c}}-v)\sum_{s=-\infty }^{\infty
}\delta (t-sT) .  \label{HN}
\end{equation}
Since $y_{{\rm c}}$ does not appear in (\ref{HN}), its conjugate mate $x_{
{\rm c}}$ is {\em conserved} and can therefore be treated as a parameter.
This reduces (\ref{H}) to an ensemble of periodically kicked harmonic
oscillators parametrized by $x_{{\rm c}}$. Classically, these degenerate
systems are expected to exhibit unbounded chaotic diffusion in the $(u,\ v)$
phase plane for arbitrarily small values of $K$, especially under resonance
conditions, with $\alpha \equiv \omega T$ a rational multiple of $2\pi $ 
\cite{zas,da}; this diffusion is observed to take place on a ``stochastic
web'' whose symmetry depends on $\alpha $. Examples of square-symmetry
webs ($\alpha =\pi /2$) are shown Fig. 1. The KHM is exactly related to (\ref
{HN}), in essence, only for $\alpha =\pi /2$ and for {\em very special}
values of $x_{{\rm c}}$, $x_{{\rm c}}=0,\ \pi $ (see Sec. A of Appendix and
generalizations of this relation in Refs. \cite{d,d1}). Now, the classical
properties of (\ref{HN}), in particular the structure of the stochastic web,
its width, and the diffusion rate on it, are known to depend {\em strongly}
on $x_{{\rm c}}$ \cite{da,pr} (see also Sec. III). In addition, one should
notice that {\em realistic} classical ensembles or 2D wave packets for (\ref
{H}) exhibit {\em all} values of $x_{{\rm c}}$. These facts motivate the
investigation of the quantum properties of (\ref{HN}) for arbitrary $x_{{\rm
c}}$. Except of general results \cite{d,d1}, these properties appear to be
essentially unexplored.\newline

In this paper, we present a first study of the quantized system
(\ref{HN}) for general $x_{{\rm c}}$. We focus on the weak-chaos regime
of small $K$ and, as in the classical studies \cite{da,pr}, on the case
of $\alpha =\pi /2 $. The global spectral features of the system at
fixed $x_{\rm c}$ are exhibited by a suitably defined plot of the
quasienergy spectrum as a function of a scaled Planck constant $\hbar$.
The term ``Hofstadter butterfly'' (HB) is extended so as to refer to
such a plot. The quantum properties are found to depend sensitively on
$x_{\rm c}$.  Most significant changes take place when $x_{{\rm c}}$
approaches the value of $\pi /2$ at which quantum antiresonance (QAR),
i.e., exactly periodic recurrences \cite{d,des}, occurs for integer
$\hbar /(2\pi )$. While the HB for most values of $x_{{\rm c}}$ is
approximately the standard one of the Harper model \cite{drh}, the HB
for $x_{{\rm c}}=\pi /2$ is, up to a local scaling, a perturbed
``doubled'' version of the standard HB. This change of the HB structure
in the ``QAR transition'' $x_{{\rm c}}\rightarrow \pi /2$ is explained
on the basis of an effective Hamiltonian which is derived as a power
expansion in a small parameter. For irrational $\hbar /(2\pi )$, the
time evolution of wave packets is observed to exhibit approximately an
asymptotic quantum diffusion. This is characterized by either a
diffusion coefficient $D(x_{{\rm c}})$ at fixed $x_{{\rm c}}$ or a
global coefficient $\overline{D}$, given by a weighted average of
$D(x_{{\rm c}})$ over $x_{{\rm c}}$. An approximate formula for
$D(x_{{\rm c}})$ in a simple class of wave packets is derived from the
effective Hamiltonian and is verified numerically. As $x_{{\rm
c}}\rightarrow \pi /2$, $D(x_{ {\rm c}})$ decreases and QAR is
manifested by the fact that $D(\pi /2)$ is significantly smaller than
typical values of $D(x_{{\rm c}})$. While these phenomena are of a
purely quantum nature, they have classical analogs. The paper is
organized as follows. In Sec. II, we recall the QAR phenomenon \cite{d,des}
and study general spectral properties of the system (\ref{HN}), also on
the basis of an effective Hamiltonian which is derived as an expansion
in a small parameter. In Sec. III, we consider classical limits and
analogs of some of the results in Sec. II. In Sec. IV, we study the
quantum diffusion of wave packets for general $x_{{\rm c}}$, using also
the effective-Hamiltonian approximation. A summary and conclusions are
presented in Sec. V. Proofs of several statements in Secs. I, II, and IV
are given in the Appendix.\newline

\begin{center}
{\bf II. QUANTUM ANTIRESONANCE, GENERAL SPECTRAL PROPERTIES,}

{\bf AND EFFECTIVE HAMILTONIAN}\\[0pt]
\end{center}

To quantize the system (\ref{HN}), $(u,\ v)$ are replaced by the
corresponding conjugate operators $(\hat{u},\ \hat{v})$. From the
definitions $u=\Pi _{x}/\omega ,\ v=\Pi _{y}/\omega $, where $(\Pi
_{x},\ \Pi _{y})={\bf p}-{\bf B\times r}/(2c)$ and $\omega =B/c$, it is
easy to see that $(\hat{u},\ \hat{v})$ satisfy $[\hat{u},\
\hat{v}]=i\hbar /\omega$. We choose units such that $\omega =1$,  
so that $[\hat{u},\ \hat{v} ]=i\hbar$, where $\hbar$ is a scaled
Planck constant. The one-period evolution operator for the quantized
system (\ref{HN}) will be denoted by $\hat{U}_{T}(x_{{\rm c}})$. Let us
now recall the phenomenon of {\em quantum antiresonance} (QAR) \cite{d}.
Under classical resonance conditions, $\alpha =\omega T=2\pi j/l$ ($j$
and $l$ are coprime integers), QAR for the system (\ref{HN}) is defined
by $\hat{U}_{T}^{l}(x_{{\rm c}})=$ constant phase factor, i.e., exactly
periodic recurrences with the natural resonance period $lT$. As shown in
Ref. \cite{d}, QAR occurs in the nonintegrable ($l>2$) case of a general
kicked harmonic oscillator only if three conditions are satisfied: (a)
The kicking periodic potential is an {\em odd} function, up to an additive
constant; this implies that $x_{{\rm c}}=\pi /2,\ 3\pi /2$ for the
potential $\cos (x_{{\rm c}}-v)$ in Eq. (\ref{HN}). (b) $l=4$ or $l=6$,
corresponding to stochastic webs with square or hexagonal symmetry,
respectively. (c) $\hbar /(2\pi )$ is integer for $l=4$ while
$\sqrt{3}\hbar /(4\pi )$ is integer for $l=6$. When all these conditions
hold, it follows from the results in Ref. \cite{d} that
$\hat{U}_{T}^{l}(x_{{\rm c}})=-1$ for $x_{{\rm c}}=\pi /2,\ 3\pi /2$ and
$l=4,\ 6$. In this paper, only the case of $l=4$ with $j/l=1/4$ ($\alpha
=\pi /2$) will be considered. Using general results \cite{d}, one finds
that the evolution operator $\hat{U}_{T}^{4}(x_{{\rm c}})$ in this case
is equal to $-\hat{U}(x_{{\rm c}})$, where
\begin{equation}
\hat{U}(x_{{\rm c}})=\exp [i\mu \cos (x_{{\rm c}}-\hat{u})]\exp [i\mu \cos
(x_{{\rm c}}+\hat{v})]\exp [i\mu \cos (x_{{\rm c}}+\hat{u})]\exp [i\mu \cos
(x_{{\rm c}}-\hat{v})]  \label{U4}
\end{equation}
and $\mu =K/\hbar $. We show in the Appendix (Sec. A) that the range of $x_{
{\rm c}}$ in Eq. (\ref{U4}) can be restricted to $[0,\ \pi /2]$ without
loss of generality. Thus, only the QAR point $x_{{\rm c}}=\pi /2$ will
be considered; $\hat{U}(\pi /2)=1$ for integer $\hbar /(2\pi )$.\newline

We now extend the term ``Hofstadter butterfly'' (HB) \cite{drh} to the
operator (\ref{U4}). Let us denote by $E$ and $\Psi $, respectively, the
quasienergies (QEs) and QE eigenstates, defined by $\hat{U}(x_{{\rm
c}})\Psi =\exp (-i\mu E)\Psi $, with $E$ lying in the interval $[-\pi
/\mu ,\ \pi /\mu )$. We then define the HB as the plot of the QE spectrum
$E(\hbar )$ as a function of $\hbar $, for $0\leq \hbar \leq 2\pi $, at
{\em fixed} $\mu $. Such a plot was introduced in Ref. \cite{gkp1} for
the KHM and is motivated by the fact, shown in the Appendix (Sec. B), that 
$E(\hbar )$ is $2\pi$-periodic in $\hbar $ at fixed $\mu $. A small value of 
$\mu $ characterizes a HB associated with a ``semiclassical weak-chaos
regime'', i.e., a regime of small $K=\hbar \mu$ for all small values of
$\hbar $, $\hbar <1$. We also show in the Appendix (Sec. B) that: (a)
For rational $\hbar /(2\pi )=q/p$ ($q$ and $p$ are coprime integers), the
QE spectrum consists of $p$ bands, each $q$-fold degenerate. (b) The HB
has reflection symmetry around $\hbar =\pi $ [i.e., $E(2\pi -\hbar
)=E(\hbar )$] only if $x_{{\rm c}}=0$ or $x_{{\rm c}}=\pi /2$. Examples
of HBs, plotted for all $\hbar =2\pi q/p$ with $1\leq q\leq p\leq 50 $,
are shown in Fig. 2 for three values of $x_{{\rm c}}$. We see that for
$x_{{\rm c}}=\pi /2$ (Fig. 2(c)) the QE spectrum $E\rightarrow 0$ as
$\hbar \rightarrow 2\pi $ or $\hbar \rightarrow 0$, reflecting the QAR
phenomenon, $\hat{U}(\pi /2)=1$.\newline

The unitary operator (\ref{U4}) can be formally written as
$\hat{U}(x_{{\rm c }})=\exp [-i\mu \hat{H}_{{\rm eff}}(x_{{\rm c}})]$,
where $\hat{H}_{{\rm eff}}(x_{{\rm c}})$ is an effective Hamiltonian for
the problem. Clearly, if $E'$ is the ``energy''spectrum of $\hat{H}_{{\rm 
eff}}(x_{{\rm c}})$, the QE spectrum $E$ is just $E'$ modulo the interval
$[-\pi /\mu ,\ \pi /\mu )$. Thus, the ``unfolded'' HB is the spectrum $E'$ 
plotted as a function of $\hbar$ ($0\leq \hbar \leq 2\pi$) at fixed
$\mu$. The dependence of some general HB features on $x_c$ can be
understood on the basis of an expression for $\hat{H}_{{\rm eff}}(x_{{\rm
c}})$ which we now derive. We first note that the arguments of the
exponents in Eq. (\ref{U4}) involve simple operators $\exp (\pm
i\hat{u})$ and $\exp (\pm i\hat{v})$, giving just translations by $\pm
\hbar $ in the $(u,\ v)$ phase plane. Using then the known formula 
\cite{rmw} 
\begin{equation}
\exp (A)\exp (B)=\exp \left( A+B+\frac{1}{2}[A,\ B]+\frac{1}{12}[A,\ [A,\
B]]+\frac{1}{12}[[A,\ B],\ B]+\dots \right) ,  \label{AB}
\end{equation}
valid for arbitrary operators $A$ and $B$, one can verify by tedious but
straightforward algebra that $\hat{H}_{{\rm eff}}(x_{{\rm c}})$ can be
expressed as a power expansion in a parameter $\epsilon $, 
\begin{equation}
\hat{H}_{{\rm eff}}(x_{{\rm c}})=\sum_{r=0}^{\infty }\epsilon ^{r}\hat{H}
_{r}(x_{{\rm c}}),\ \ \ \ \ \ \epsilon =\mu \sin (\hbar /2)=\frac{K}{2}\frac{
\sin (\hbar /2)}{\hbar /2}.  \label{He}
\end{equation}
The zero-order coefficient in Eq. (\ref{He}) is 
\begin{equation}
\hat{H}_{0}(x_{{\rm c}})=-2\cos (x_{{\rm c}})[\cos (\hat{u})+\cos (\hat{v})].
\label{H0}
\end{equation}
The operator (\ref{H0}) is a Harper Hamiltonian and if $\epsilon $ is
sufficiently small, $2|\cos (x_{{\rm c}})/\epsilon |\gg 1$, is the
dominant term in the expansion (\ref{He}) for most values of $x_{{\rm c}}$. 
For general $\mu$, $\epsilon $ is small provided $\hbar /(2\pi )$ is
sufficiently close to an integer. In a weak-chaos (small $\mu $ or $K$)
regime, on which we shall focus, $\epsilon $ is globally small for {\em all} 
$\hbar $. Then, for $2|\cos (x_{{\rm c}})/\epsilon |\gg 1$, the HB is
essentially the standard one for the Harper model \cite{drh}, see Fig. 2(a).
For $2|\cos (x_{{\rm c}})/\epsilon |\lesssim 1$, higher-order coefficients
in Eq. (\ref{He}) become significant, especially for $\hbar \approx \pi$
(largest $\epsilon $ at fixed $\mu$), leading to a clearly non-standard HB
with a visible reflection-symmetry breaking; see Fig. 2(b).\newline

For $x_{{\rm c}}=\pi /2$, $\hat{H}_{0}$ {\em vanishes}. If also $\hbar
/(2\pi )$ is integer, $\epsilon =0$ and thus $\hat{H}_{{\rm eff}}=0$,
implying $\hat{U}(\pi /2)=1$ (QAR). Small values of $\epsilon $ for $x_{{\rm
c}}=\pi /2$ define a {\em QAR vicinity}; thus, a weak-chaos regime at $x_{
{\rm c}}=\pi /2$ corresponds to a QAR vicinity for arbitrary $\hbar $. Now,
$\hat{H}_{{\rm eff}}(\pi /2)=\epsilon \hat{H}_{1}+\epsilon ^{2}\hat{H}
_{2}+\dots $, and we find that 
\begin{equation}
\hat{H}_{1}(\pi /2)=-[\cos (\hat{u}+\hat{v})+\cos (\hat{u}-\hat{v})].
\label{H1s}
\end{equation}
Since the operators $\hat{u}^{\prime }=\hat{u}+\hat{v}$ and $\hat{v}^{\prime
}=\hat{v}-\hat{u}$ are clearly conjugate, $[\hat{u}^{\prime },\ \hat{v}
^{\prime }]=i\hbar ^{\prime }$ with $\hbar ^{\prime }=2\hbar $, we see that
$\hat{H}_{1}$ in Eq. (\ref{H1s}) is again a Harper Hamiltonian. When
$\hbar $ varies from $0$ to $2\pi $, $\hbar ^{\prime }=2\hbar $ varies
from $0$ to $4\pi $. Then, the HB for $\hat{H}_{{\rm eff}}/\epsilon
=\hat{H}_{1}+\epsilon \hat{H}_{2}+\dots $ is a perturbed doubled version of 
the standard HB, see Fig. 3. A simple measure of the perturbation of this
HB is $P=|1-\Delta E^{\prime }(\pi )/\Delta E^{\prime }(2\pi )|$, where
$\Delta E^{\prime}(\hbar )$ is the width of the spectrum of $\hat{H}_{{\rm 
eff}}/\epsilon $. The exact result $P=1-\mu ^{-2}\arcsin [\sin ^{2}(\mu
)]=\mu ^{2}/3+\dots $ is easily derived (see note \cite{note}). As $\hbar
$ approaches its QAR value of $2\pi $ (or $0$ at fixed $\mu $), the width
$\Delta E(\hbar )\approx \mu \sin (\hbar /2)\Delta E^{\prime }(2\pi )$ of
the QE spectrum vanishes almost linearly in $\hbar $, see Fig. 2(c).\newline

A simple consequence of the approximately doubled structure of the HB for $
\hat{H}_{{\rm eff}}(\pi /2)/\epsilon $ is that if $\hbar =2\pi q/p$ and $p$
is {\em even}, $p=2p^{\prime }$, the $p$ bands in this HB actually form $
p^{\prime }$ pairs of overlapping bands. This is because the spectrum of the
Harper Hamiltonian (\ref{H1s}) for $\hbar ^{\prime }=2\hbar =2\pi
q/p^{\prime }$ always consists of $p^{\prime }$ nonoverlapping bands \cite
{drh}. Each such band must then correspond to a pair of overlapping bands of 
$\hat{H}_{{\rm eff}}$ for sufficiently small $\epsilon $. Thus, the number
of gaps in the spectrum for even $p$\ is more than halved when $x_{{\rm c}}$
is varied from $0$ to $\pi /2$.\newline

\begin{center}
{\bf III. CLASSICAL LIMITS AND ANALOGS}\\[0pt]
\end{center}

In this section, we consider classical limits and analogs of some of the
results above. The classical one-period map for the system (\ref{HN}) in
the case of $\alpha =\pi /2$ is \cite{da}: 
\begin{equation}
M:\ \ u_{s+1}=v_{s},\ \ \ \ v_{s+1}=-u_{s}+K\sin (x_{{\rm c}}-v_{s}),
\label{M}
\end{equation}
The fourth iterate $M^{4}$ of $M$ is the classical map corresponding to
the evolution operator $\hat{U} (x_{{\rm c}})$ in Eq. (\ref{U4}). One
can easily show that $M^{4}$ is a near-identity map for small $K$
\cite{pr}, i.e., $u_{t+1}=u_{t}+O(K)$, $v_{t+1}=v_{t}+O(K)$, where $t$
denotes now an integer time index counting iterations of $M^{4}$. It
is then clear from the expression $\hat{U} (x_{{\rm c}})=\exp
[-iK\hat{H}_{{\rm eff}}(x_{{\rm c}})/\hbar ]$ that the classical limit
($\hbar \rightarrow 0$, $\epsilon \rightarrow K/2$) of $\hat{ H}_{{\rm
eff}}(x_{{\rm c}})$ is an integrable Hamiltonian $H_{{\rm eff}}(u,\ v;\
x_{{\rm c}})$ generating, in a continuous time denoted here by
$t^{\prime }$, a phase-space flow which approximates the near-integrable
map $M^{4}$ at times $t^{\prime }=Kt$:
\begin{equation}
\frac{\partial H_{{\rm eff}}}{\partial v}=\frac{du}{dt^{\prime }}\approx 
\frac{u_{t+1}-u_{t}}{K},\ \ \ \ -\frac{\partial H_{{\rm eff}}}{\partial u}=
\frac{dv}{dt^{\prime }}\approx \frac{v_{t+1}-v_{t}}{K}.  \label{Hf}
\end{equation}
Expressions for $KH_{{\rm eff}}/4$ as power expansions in $K$\ are given in
Ref. \cite{pr}. Like $\hat{U}(x_{{\rm c}})$, $M^{4}$ and $H_{{\rm eff}}(u,\
v;\ x_{{\rm c}})$ are generally periodic in the phase space $(u,\ v)$ with a
unit cell of area $4\pi ^{2}$, so that the orbit structure of $M^{4}$ or the
flow (\ref{Hf}) feature the same periodicity, as clearly shown in Figs. 1(a)
and 1(b). For $x_{{\rm c}}=\pi /2$, however, the dominant term in $H_{{\rm
eff}}(u,\ v;\ \pi /2)$ is the classical analog $H_{1}$ of (\ref{H1s}) which
is periodic with a unit cell of area $2\pi ^{2}$, {\em half} the size of the
usual unit cell; this periodicity is then approximately exhibited by the
corresponding orbit structure, compare Fig. 1(c) with Fig. 1(a). This is the
classical fingerprint in the approximate doubling of the HB for $x_{{\rm c}
}=\pi /2$ (see previous section and Fig. 3).\newline

The hyperbolic fixed points of $M^{4}$ form a periodic array connected
by the stochastic web, see Fig. 1. As shown in Ref. \cite{pr}, the width
of the web for small $K$ is proportional to $\exp [-\pi ^{2}/\ln
(\lambda )]$, where $\lambda $ is the largest eigenvalue of the
linearization of $M^{4}$ at a hyperbolic fixed point and is given by
$\lambda \approx 1+2K\cos (x_{{\rm c }})$ for $\cos (x_{{\rm c}})\gg
K/4$ and by $\lambda \approx 1+K^{2}$ for $x_{{\rm c}}=\pi /2$. Thus,
the latter case is characterized by a significantly small web width and
slow chaotic diffusion relative to those for most values of $x_{{\rm
c}}$ \cite{pr}. These features may be viewed as classical analogs of
the QAR phenomenon, including some quantum behaviors in a QAR vicinity
(see Sec. IV C and Sec. V).\newline

\begin{center}
{\bf IV. QUANTUM DIFFUSION FOR GENERAL }$x_{{\rm c}}$\\[0pt]
\end{center}

In this section, we study the quantum diffusion exhibited by wave packets
evolving under the operator (\ref{U4}) and its effective-Hamiltonian
approximations.

\begin{center}
{\bf A. General quantum evolution and periodic wave packets}\\[0pt]
\end{center}

A general wave packet $|\Phi \rangle $ for the original 2D problem (\ref{H})
can be naturally expressed in a representation based on the two degrees of
freedom (conjugate pairs) $(\hat{u},\ \hat{v})$ and $(\hat{x}_{{\rm c}},\ 
\hat{y}_{{\rm c}})$, for example, in the $(u,\ x_{{\rm c}})$ representation, 
$\left\langle u,\ x_{{\rm c}}|\Phi \right\rangle =\Phi (u,\ x_{{\rm c}})$.
The time evolution of $\Phi (u,\ x_{{\rm c}})$ can be decomposed into the
independent evolutions for the separate, conserved $x_{{\rm c}}$ values, 
\begin{equation}
\Phi (u,\ x_{{\rm c}};\ t)=\hat{U}^{t}(x_{{\rm c}})\Phi (u,\ x_{{\rm c}}),
\label{exc}
\end{equation}
$t$ being the integer time index. We show in the Appendix (Sec. C) that the
quantum dynamics (\ref{exc}) can be fully reproduced from that of wave
packets on a cylindrical phase space, $0\leq u<2\pi $, $-\infty <v<\infty $,
with $\Phi (u,\ x_{{\rm c}})$ $2\pi $-periodic in $u$: 
\begin{equation}
\Phi (u,\ x_{{\rm c}})=\sum_{n=-\infty }^{\infty }\varphi (n,\ x_{{\rm c}
})\exp (inu).  \label{pwp}
\end{equation}
Here the Fourier coefficients $\varphi (n,\ x_{{\rm c}})$ give the $(v,\ x_{
{\rm c}})$ representation of $|\Phi \rangle $ on the cylinder, with $v$\
quantized in units of $\hbar $, $v=n\hbar $. The quantum evolution of the
wave packet (\ref{pwp}) in the $(n,\ x_{{\rm c}})$ representation, 
\begin{equation}
\varphi (n,\ x_{{\rm c}};\ t)=\hat{U}^{t}(x_{{\rm c}})\varphi (n,\ x_{{\rm c}
}),  \label{enxc}
\end{equation}
can be easily calculated as in the kicked-rotor case \cite{kr}, by using a
fast Fourier transform to switch from the position ($u$) to the momentum 
($v=n\hbar$) representation (and vice-versa) before the application of each
factor in Eq. (\ref{U4}). The spreading of the wave packet (\ref{enxc})
along the cylinder at fixed $x_{{\rm c}}$ is measured by the expectation
value $\left\langle \hat{v}^{2}\right\rangle _{t}$ of $\hat{v}^{2}$ in 
$\varphi (n,\ x_{{\rm c}};\ t)$: 
\begin{equation}
\left\langle \hat{v}^{2}\right\rangle _{t}=\frac{\hbar ^{2}}{N(x_{{\rm c}})}
\sum_{n=-\infty }^{\infty }n^{2}|\varphi (n,\ x_{{\rm c}};\ t)|^{2},\ \ \ \
\ \ N(x_{{\rm c}})=\sum_{n=-\infty }^{\infty }|\varphi (n,\ x_{{\rm c}
})|^{2}.  \label{ev2n}
\end{equation}
The expectation value of $\hat{v}^{2}$ in the original {\em 2D wave packet}
$\varphi (n,\ x_{{\rm c}};\ t)$ (both $n$ and $x_{{\rm c}}$ are variables) 
is given by 
\begin{equation}
\overline{\left\langle \hat{v}^{2}\right\rangle _{t}}=\hbar
^{2}\sum_{n=-\infty }^{\infty }n^{2}\int |\varphi (n,\ x_{{\rm c}};\
t)|^{2}dx_{{\rm c}},  \label{ev2ng}
\end{equation}
where normalization of $\varphi $ over $(n,\ x_{{\rm c}})$ is assumed, $\int
N(x_{{\rm c}})dx_{{\rm c}}=1$. We also show in the Appendix (Sec. C) that
our main results below concerning $\left\langle \hat{v}^{2}\right\rangle
_{t} $ can be extended to the expectation value $\left\langle \omega ^{2}(
\hat{u}^{2}+\hat{v}^{2})/2\right\rangle _{t}$ of the kinetic energy in an
arbitrary wave packet (\ref{exc}). Thus, for convenience and without loss of
generality, we shall restrict our attention to periodic wave packets 
(\ref{pwp}) and to the expectation values (\ref{ev2n}) and (\ref{ev2ng}).

\begin{center}
{\bf B. Quantum diffusion at fixed }$x_{{\rm c}}${\bf \ and global quantum
diffusion}\\[0pt]
\end{center}

We have seen in Sec. II that if $\epsilon $ is sufficiently small the
dominant term in the expansion (\ref{He}) of $\hat{H}_{{\rm eff}}(x_{{\rm c}
})$ for most values of $x_{{\rm c}}$, including $x_{{\rm c}}=\pi /2$, is a
Harper Hamiltonian. It is known \cite{gkp,wa} that for irrational $\hbar
/(2\pi )$ the latter Hamiltonian exhibits a fractal spectrum and,
asymptotically in time, an approximate quantum diffusion of the second
moment of a wave packet. We can therefore expect a similar behavior of $
\left\langle \hat{v}^{2}\right\rangle _{t}$ under $\hat{U}(x_{{\rm c}})=\exp
[-i\mu \hat{H}_{{\rm eff}}(x_{{\rm c}})]$ for irrational $\hbar /(2\pi )$
and sufficiently large $t$: $\left\langle \hat{v}^{2}\right\rangle
_{t}\approx 2D(x_{{\rm c}})t$, where $D(x_{{\rm c}})$ is the diffusion
coefficient at fixed $x_{{\rm c}}$.\newline

As an illustration, let us choose the initial wave packet (\ref{pwp}) as a
periodized coherent state centered on some $x_{{\rm c}}$-dependent point 
$(\overline{u},\ \overline{v})$, 
\begin{equation}
\Phi (u,\ x_{{\rm c}})=\left( \pi \hbar \right) ^{-1/4}\sum_{m=-\infty
}^{\infty }\exp \left\{ i\frac{\overline{v}(x_{{\rm c}})(u-2\pi m)}{\hbar }-
\frac{[u-2\pi m-\overline{u}(x_{{\rm c}})]^{2}}{2\hbar }\right\} .
\label{pcs}
\end{equation}
For sufficiently small $\hbar $, a classical quantity analogous to the
expectation value $\left\langle \hat{v}^{2}\right\rangle _{t}$ in the
wave packet (\ref{pcs}) is the average $\left\langle
v_{t}^{2}\right\rangle $ of $v_{t}^{2}$ over an ensemble of initial
conditions $(u_{0},\ v_{0})$ uniformly distributed in a disk centered on
$(\overline{u},\ \overline{v})$ and of radius $\sqrt{2\hbar }$;
$(u_{t},\ v_{t})$ is the $t$th iterate of $(u_{0},\ v_{0})$ under the
map $M^{4}$ (see previous section). Fig. 4 shows log-log plots of
$\left\langle \hat{v}^{2}\right\rangle _{t}$ and $\left\langle
v_{t}^{2}\right\rangle $ for $\hbar /(2\pi )=[51+(\sqrt{5}-1)/2]^{-1}$,
$K=0.157$, and $0<t\leq 10^{5}$ in the three cases of $x_{{\rm c}}$
considered in Figs. 1 and 2; in each case, $(\overline{u},\
\overline{v})$ was chosen as a hyperbolic fixed point of the
corresponding map $M^{4}$. We see that $\left\langle \hat{v}
^{2}\right\rangle _{t}$ and $\left\langle v_{t}^{2}\right\rangle $
essentially coincide up to some time $t=t^{\ast }(x_{{\rm c}})$, after
which $\left\langle v_{t}^{2}\right\rangle $ stops spreading and there
occurs a crossover of $\left\langle \hat{v}^{2}\right\rangle _{t}$ to an
approximate quantum diffusion, $\left\langle \hat{v}^{2}\right\rangle
_{t}\approx 2D(x_{ {\rm c}})t$. In the case of $x_{{\rm c}}=\pi /2$,
this diffusion is shown more clearly by the normal plot of $\left\langle
\hat{v}^{2}\right\rangle _{t}$ in the inset of Fig. 4.\newline

These results can be understood as follows. The irrational value of $\hbar
/(2\pi )$ chosen above represents a semiclassical regime in which $h=2\pi
\hbar $ is small relative to the area $4\pi ^{2}$ of a unit cell but large
relative to the width $\sim \exp [-\pi ^{2}/\ln (\lambda )]$ of the
stochastic web for $K=0.157$ (see classical details in the previous
section). Thus, classical chaos leaves almost no fingerprints in the quantum
properties which should then resemble those of the integrable approximation
of the system given by the effective Hamiltonian $H_{{\rm eff}}(x_{{\rm c}})$,
with the stochastic web replaced by its separatrix ``skeleton'' \cite
{zas,pr}. A classical ensemble of width $\sim \sqrt{2\hbar }$ centered on a
hyperbolic fixed point will lie mostly in the stable regions within the unit
cells and will spread to its maximal extent (of the order of the size of one
unit cell) in a time $t=t^{\ast }(x_{{\rm c}})$ roughly proportional to $\ln
(1/\hbar )/\ln [\lambda (x_{{\rm c}})]$; this agrees with numerical
estimates of $t^{\ast }(x_{{\rm c}})$ in all cases considered, including
those shown in Fig. 4. The quantum wave packet, however, will continue to
spread due to tunnelling between neighboring unit cells, leading to quantum
diffusion \cite{gkp,wa}.\newline

Assuming the asymptotic quantum-diffusion behavior of $\left\langle\hat{v}
^{2}\right\rangle _{t}$, $\left\langle \hat{v} ^{2}\right\rangle _{t}\approx 
2D(x_{{\rm c}})t$, to hold approximately for all $x_{{\rm c}}$, one finds
from Eqs. (\ref{ev2n}) and (\ref{ev2ng}) that the 2D wave packet (\ref{enxc}) 
can be characterized by a more realistic quantity, its global diffusion
coefficient $\overline{D}$, given by a weighted average of $D(x_{{\rm
c}})$ over $x_{{\rm c}}$: 
\begin{equation}
\overline{\left\langle \hat{v}^{2}\right\rangle _{t}}\approx 2\overline{D}
t,\ \ \ \ \ \ \overline{D}=\int N(x_{{\rm c}})D(x_{{\rm c}})dx_{{\rm c}},
\label{gad}
\end{equation}
where $N(x_{{\rm c}})$ is defined in Eq. (\ref{ev2n}). The global diffusion
(\ref{gad}) will be briefly considered at the end of this section.

\begin{center}
{\bf C. }$x_{{\rm c}}${\bf -dependence of the quantum-diffusion coefficient
for simple wave packets}\\[0pt]
\end{center}

We now consider the $x_{{\rm c}}$-dependence of the quantum-diffusion
coefficient $D$. It is clear from Eq. (\ref{enxc}) that this dependence is 
generally due to that of both $\hat{U}(x_{{\rm c}})$ and the initial wave
packet $\varphi (n,\ x_{{\rm c}})$. For simplicity, we shall restrict
ourselves from now on to the class of wave packets for which one has a
separation of the $(n,\ x_{{\rm c}})$ variables, $\varphi (n,\ x_{{\rm c}
})=\chi (x_{{\rm c}})\psi (n)$ [$\chi (x_{{\rm c}})$ and $\psi (n)$ are
assumed to be normalized]. Then, $\varphi (n,\ x_{{\rm c}};\ t)=\chi (x_{
{\rm c}})\hat{U}^{t}(x_{{\rm c}})\psi (n)$, so that $\left\langle \hat{v}
^{2}\right\rangle _{t}$ in Eq. (\ref{ev2n}) is independent of $\chi
(x_{{\rm c}})$ and the $x_{{\rm c}}$-dependence of $\left\langle \hat{v}
^{2}\right\rangle _{t}$ for fixed $\psi (n)$ is completely due to that of 
$\hat{U}(x_{{\rm c}})$. This allows one to derive an approximate formula for 
$D(x_{{\rm c}})$ in the case of $2|\cos (x_{{\rm c}})/\epsilon |\gg 1$. In
this case, which covers most values of $x_{{\rm c}}$, the effective
Hamiltonian $\hat{H}_{{\rm eff}}\approx \hat{H}_{0}$, where $\hat{H}_{0}$ is
given by Eq. (\ref{H0}). The $t$th power of the approximate evolution
operator $\hat{U}_{0}=\exp (-i\mu \hat{H}_{0})$ can be written as follows: 
\begin{equation}
\hat{U}_{0}^{t}(x_{{\rm c}})=\exp \left\{ 2iK\cos (x_{{\rm c}})t\left[ \cos
(\hat{u})+\cos (\hat{v})\right] /\hbar \right\} .  \label{U0t}
\end{equation}
Rel. (\ref{U0t}) implies that the wave packet $\psi (n,\ x_{{\rm c}};\ t)=
\hat{U}_{0}^{t}(x_{{\rm c}})\psi (n)$ depends on $K$, $x_{{\rm c}}$, and $t$
only through the combination $K\cos (x_{{\rm c}})t$. Thus, if one assumes an
asymptotic quantum-diffusive behavior, $\left\langle \hat{v}
^{2}\right\rangle _{t}\approx 2D(x_{{\rm c}})t$, accurately described by 
$\hat{U}_{0}$, $D(x_{{\rm c}})$ must be approximately linear in $K\cos 
(x_{{\rm c}})$: 
\begin{equation}
D(x_{{\rm c}})\approx 2D_{{\rm H}}^{(\pm )}K\cos (x_{{\rm c}}),\ \ \ \ \ \ \
2|\cos (x_{{\rm c}})/\epsilon |\gg 1.  \label{Dxc}
\end{equation}
Here the proportionality constants $D_{{\rm H}}^{(+)}$ and $D_{{\rm H}
}^{(-)} $ are associated with the $x_{{\rm c}}$-intervals where $\cos (x_{
{\rm c}})>0 $ and $\cos (x_{{\rm c}})<0$, respectively; assuming that $K>0$
for definiteness, one has $D_{{\rm H}}^{(+)}>0$ and $D_{{\rm H}}^{(-)}<0$
since $D(x_{{\rm c}})>0$. The constants $D_{{\rm H}}^{(+)}$ and $|D_{{\rm 
H}}^{(-)}| $ are just the diffusion coefficients for $\left\langle \hat{v}
^{2}\right\rangle _{t}$ under the standard Harper evolution operator $\hat{U}
_{{\rm H}}=\exp \left\{ i\left[ \cos (\hat{u})+\cos (\hat{v})\right] /\hbar
\right\} $ and its inverse $\hat{U}_{{\rm H}}^{-1}$, respectively. For
general $\psi (n)$, $D_{{\rm H}}^{(+)}\neq |D_{{\rm H}}^{(-)}|$. However, if 
$\psi (n)$ satisfies some symmetry properties, e.g., its $u$ representation
is given by the right-hand side of Eq. (\ref{pcs}) with $\overline{u}$ a
multiple of $\pi $ (as in the example below), one can show that $D_{{\rm H}}
^{(+)}=|D_{{\rm H}}^{(-)}|$ (see Appendix, Sec. D). Formula (\ref{Dxc}) can
then be rewritten as $D(x_{{\rm c}})\approx 2D_{{\rm H}}K|\cos (x_{{\rm c}
})| $ with the same constant $D_{{\rm H}}$ for all $x_{{\rm c}}$ satisfying 
$2|\cos (x_{{\rm c}})/\epsilon |\gg 1$.\newline

As a check of formula (\ref{Dxc}), we have calculated $D(x_{{\rm c}})$ for 
$\hbar /(2\pi )=[51+(\sqrt{5}-1)/2]^{-1}$, $K=0.157$, and $x_{{\rm c}}=k\pi
/40$, $k=0,\ 1,\dots ,\ 18$, by a linear fit to $\left\langle \hat{v}
^{2}\right\rangle _{t}$ ($0<t\leq 10^{4}$), evolving under the exact
operator $\hat{U}(x_{{\rm c}})$. The $u$ representation of the initial wave
packet $\psi (n)$ is given by the right-hand side of Eq. (\ref{pcs}) with 
$(\overline{u},\ \overline{v})=(\pi ,\ 0)$, a hyperbolic fixed point for $x_{
{\rm c}}=0$ (see Fig. 1(a)). The results, shown in Fig. 5, agree
satisfactorily with a least-square fit of the function (\ref{Dxc}) to the
data.\newline

Consider now the case of $x_{{\rm c}}=\pi /2$. From the fact that $\hat{H}_{
{\rm eff}}(\pi /2)\approx \epsilon \hat{H}_{1}(\pi /2)=O(K)$, one expects,
using arguments similar to those leading to formula (\ref{Dxc}), that $D(\pi
/2)$ should be approximately quadratic in $K$. In Table I, we present
results concerning the $K$-dependence of $D(0)/K$ and $D(\pi /2)/K^{2}$. The
values of $D(0)$ and $D(\pi /2)$ were calculated by a linear fit to $
\left\langle \hat{v}^{2}\right\rangle _{t}$ ($0<t\leq 10^{4}$ for $x_{{\rm c}
}=0$ and $0<t\leq 10^{5}$ for $x_{{\rm c}}=\pi /2$), evolving under the
exact operator (\ref{U4}); for both $x_{{\rm c}}=0$ and $x_{{\rm c}}=\pi /2$, 
the initial wave packet is given by the right-hand side of Eq. (\ref{pcs})
with $(\overline{u},\ \overline{v})$ chosen as a hyperbolic fixed point of
the corresponding map $M^{4}$. The results for $D(0)/K$ agree very well with
the approximate linearity in $K$ predicted by formula (\ref{Dxc}). On the
other hand, $D(\pi /2)$ appears to decrease faster than $K^{2}$ as $K$
decreases, which may indicate that high-order corrections to $\hat{H}_{{\rm
eff}}$ for $x_{{\rm c}}=\pi /2$ have a stronger impact than for $x_{{\rm c}
}=0$. In any case, Table I provides evidence that as $K$ decreases $D(\pi
/2) $ becomes significantly smaller than typical values of $D(x_{{\rm c}})$.
\newline

We end this section by a brief consideration of the global quantum diffusion
(\ref{gad}) in the case of the initial wave packets $\varphi (n,\ x_{{\rm c}
})=\chi (x_{{\rm c}})\psi (n)$. For such a wave packet, formula (\ref{Dxc})
can be used to calculate approximately the global diffusion coefficient 
$\overline{D}$, with $N(x_{{\rm c}})=|\chi (x_{{\rm c}})|^{2}$ in Eqs. 
(\ref{ev2n}) and (\ref{gad}). Choosing $N(x_{{\rm c}})$ uniform, $N(x_{{\rm c}
})=2/\pi $, for $0\leq x_{{\rm c}}<\pi /2$ and vanishing outside this
interval, one finds from Eqs. (\ref{gad}) and (\ref{Dxc}) that $\overline{D}
\approx 4D_{{\rm H}}^{(+)}K/\pi $. Much smaller values of $\overline{D}$ are
obtained if $N(x_{{\rm c}})$ is strongly localized around $x_{{\rm c}}=\pi /2
$. Fig. 6 shows the time evolution of $\overline{\left\langle \hat{v}
^{2}\right\rangle _{t}}$, given by a uniform average of $\left\langle \hat{v}
^{2}\right\rangle _{t}$ over $x_{{\rm c}}=k\pi /40$, $k=0,\ 1,\dots ,\ 18$,
where $\left\langle \hat{v}^{2}\right\rangle _{t}$ for each of these $x_{
{\rm c}}$ values was calculated as specified above for obtaining the results
in Fig. 5. We also show in Fig. 6 the functions $\left\langle \hat{v}
^{2}\right\rangle _{t}$ for $x_{{\rm c}}=0$ ($k=0$) and $x_{{\rm c}}=2\pi /5$
($k=16$). As expected, averaging over $x_{{\rm c}}$ removes from $\overline{
\left\langle \hat{v}^{2}\right\rangle _{t}}$ much of the fluctuations
exhibited by $\left\langle \hat{v}^{2}\right\rangle _{t}$ at fixed $x_{{\rm c
}}$.\newline

\begin{center}
{\bf V. SUMMARY AND CONCLUSIONS}\\[0pt]
\end{center}

Our first study of the quantized system (\ref{HN}) for general $x_{{\rm
c}}$ has revealed the significant impact of the transition $x_{{\rm c}}
\rightarrow \pi /2$ to a QAR vicinity on the spectral and quantum-diffusion
properties in a weak-chaos (small $K$) regime. In this transition, the
dominant term in the effective-Hamiltonian expansion (\ref{He}) changes from 
$\hat{H}_{0}$ to $\epsilon \hat{H}_{1}$ in a $x_{{\rm c}}$-window of width 
$\sim \epsilon $ around $x_{{\rm c}}=\pi /2$, where $\hat{H}_{0}$ vanishes.
Both $\hat{H}_{0}(x_{{\rm c}})$ and $\hat{H}_{1}(\pi /2)$ are Harper
Hamiltonians, but $\hat{H}_{1}(\pi /2)$ is periodic in phase space with a
unit cell half the size of that of $\hat{H}_{0}(x_{{\rm c}})$. As a
consequence, the HB for $x_{{\rm c}}=\pi /2$ is essentially a perturbed
doubled version of the standard HB for the Harper model (see Fig. 3). Thus,
as $x_{{\rm c}}$ is varied from $0$ to $\pi /2$, there must occur strong
changes in the HB structure and the spectral properties (see Fig. 2 and end
of Sec. II), reflecting corresponding changes in the classical web structure
due to bifurcations \cite{pr} (see Fig. 1). A more detailed investigation of
the fingerprints of these bifurcations in the quantum properties is planned
for a future work.\newline

The quantum diffusion of wave packets for the system (\ref{HN}) was
characterized by a coefficient $D(x_{{\rm c}})$ at fixed $x_{{\rm c}}$ and
by a global coefficient $\overline{D}$. Formula (\ref{Dxc}) for $D(x_{{\rm c}
})$ was derived on the basis of the approximate effective Hamiltonian $\hat{
H }_{0}(x_{{\rm c}})$ and it agrees reasonably well with numerical results
for sufficiently small $K$ (see Fig. 5 and Table I). This formula and
numerical evidence indicate that $D(x_{{\rm c}})$ is strongly reduced as $x_{
{\rm c}}\rightarrow \pi /2$: $D(\pi /2)$ appears to be not larger than 
$O(\epsilon )$ relative to typical values of $D(x_{{\rm c}})$. This
phenomenon is of a purely quantum nature but it is analogous to the
relatively slow classical chaotic diffusion for $x_{{\rm c}}=\pi /2$ in a
weak-chaos regime \cite{pr}. An interesting but apparently difficult problem
is to obtain a refined formula for $D(x_{{\rm c}})$, extending the
zero-order result (\ref{Dxc}) to the small $x_{{\rm c}}$-window around $x_{
{\rm c}}=\pi /2$ where higher-order terms in the expansion (\ref{He}) must
be taken into account.\newline

The strong variation of $D(x_{{\rm c}})$ with $x_{{\rm c}}$ is exhibited
in practice by the global quantum diffusion of a 2D wave packet for all
$x_{{\rm c}}$. If the potential $\cos (x)$ in Eq. (\ref{H}) is replaced
by $\cos (x-\theta )$, where $\theta $ is an adjustable phase, one can
``filter'' the wave-packet component associated with an arbitrary value
of $x_{{\rm c}}\approx \theta +\pi /2$\ as the component having the
smallest quantum-diffusion rate. The new quantum phenomena predicted by
the realistic, general-$x_{{\rm c}}$ approach to (\ref{H}) may be
observed in possible experimental realizations of this system.\newline

\begin{center}
{\bf ACKNOWLEDGMENTS}\\[0pt]
\end{center}

We thank M. Wilkinson for useful comments and discussions. DLD acknowledges
partial support from the Russian Ministry of Education and Science and the
US Civilian Research and Development Foundation (CRDF BRHE Program, Grants
VZ-0-010 and Y2-P-10-01).\newline

\begin{center}
{\bf APPENDIX}\\[0pt]

{\bf A. Exact relation between system (\ref{HN}) and KHM and relevant }$x_{
{\rm c}}${\bf \ range}\\[0pt]
\end{center}

The evolution operator (\ref{U4}) for the system (\ref{HN}) in the case
of $\alpha =\pi /2$ is equivalent to $\hat{U}^{\prime }(x_{{\rm
c}})=\exp \left[ -i\mu W(\hat{v})\right] \hat{U}(x_{{\rm c}})\exp \left[ 
i\mu W(\hat{v})\right] $, where $W(v)=-\cos (x_{{\rm c}}-v)$. One has 
\begin{equation}
\hat{U}^{\prime }(x_{{\rm c}})=\hat{U}_{{\rm KHM}}^{(+)}\hat{U}_{{\rm KHM}
}^{(-)},  \label{UKHM}
\end{equation}
where 
\[
\hat{U}_{{\rm KHM}}^{(\pm )}=\exp \left[ -i\mu W(\pm \hat{v})\right] \exp
\left[ -i\mu W(\pm \hat{u})\right] 
\]
is the one-period evolution operator for the generalized KHM \cite{d1,d2}
with Hamiltonian $H=W(\pm v)+W(\pm u)\sum_{s=-\infty }^{\infty }\delta
(t/K-s)$. This reduces to the usual KHM Hamiltonian, with $W(v)=\cos (v)$
(up to a sign), only for $x_{{\rm c}}=0,\ \pi $. Then, $\hat{U}_{{\rm KHM}
}^{(+)}=\hat{U}_{{\rm KHM}}^{(-)}=\hat{U}_{{\rm KHM}}$ and the exact
relation $\hat{U}^{\prime }(x_{{\rm c}})=\hat{U}_{{\rm KHM}}^{2}$ follows
from Eq. (\ref{UKHM}).\newline

The evolution operator (\ref{U4}) satisfies also the similarity relations 
\begin{equation}
\hat{U}(x_{{\rm c}}+\pi )=\hat{D}\hat{U}(x_{{\rm c}})\hat{D}^{-1},\ \ \ \hat{
U}(-x_{{\rm c}})=\hat{S}\hat{U}(x_{{\rm c}})\hat{S}^{-1},\   \label{Spim}
\end{equation}
where 
\[
\ \hat{D}=\exp (\pi i\hat{u}/\hbar )\exp (\pi i\hat{v}/\hbar ),\ \ \ \ \hat{S
}=\exp [-i\mu \cos (x_{{\rm c}}+\hat{v})]\exp [-i\mu \cos (x_{{\rm c}}-\hat{u
})]. 
\]
Since $\hat{U}(x_{{\rm c}})$ is $2\pi $-periodic in $x_{{\rm c}}$, Rels. 
(\ref{Spim}) imply that the range of $x_{{\rm c}}$ can be restricted to $[0,\
\pi /2]$ without loss of generality.

\begin{center}
{\bf B. Quasienergy band spectrum and some of its basic properties}\\[0pt]
\end{center}

Using Rel. (\ref{UKHM}), the results in Ref. \cite{d2} concerning the
quasienergy (QE) band spectra of generalized KHMs can be straightforwardly
extended to the operator $\hat{U}^{\prime }(x_{{\rm c}})$,
which is equivalent to the evolution operator $\hat{U}(x_{{\rm c}})$ in
Eq. (\ref{U4}) and thus has the same QE spectra as those of 
$\hat{U}(x_{{\rm c}})$. For $\hbar =2\pi q/p$ ($q$ and $p$ are coprime 
integers), $\hat{U}^{\prime }(x_{{\rm c}})$ and the phase-space
translations $\hat{D}_{1}=\exp (2\pi i\hat{u}/\hbar )$ and 
$\hat{D}_{2}=\exp (ip\hat{v})$ form a complete set of commuting operators.
Their simultaneous eigenstates are given, in the $v$ representation, by 
\[
\Psi _{b,{\bf w},x_{{\rm c}}}(v)=\sum_{m=0}^{p-1}\phi _{b}(m;\ {\bf w};\ x_{
{\rm c}})\sum_{n=-\infty }^{\infty }\exp [in(w_{1}+m\hbar )/q]\delta
(v-w_{2}+2\pi n/p).
\]
Here the index $b=1,...,\ p$ labels $p$ QE bands, ${\bf w}=(w_{1},\
w_{2})$ is a Bloch wave vector spanning a band and ranging in the
``Brillouin zone'' $0\leq w_{1}<\hbar $, $0\leq w_{2}<2\pi /p$, and $\left\{
\phi _{b}(m;\ {\bf w};\ x_{{\rm c}})\right\} _{m=0}^{p-1}$, $b=1,...,\ p$,
are $p$ independent vectors of coefficients. The corresponding eigenvalues
of $\hat{U}^{\prime }(x_{{\rm c}})$ are $\exp [-i\mu E_{b}({\bf w};\ x_{{\rm
c}})]$, where $E_{b}({\bf w};\ x_{{\rm c}})$ is a QE band. It is easy to see
that $\hat{U}^{\prime }(x_{{\rm c}})$ commutes with $\hat{D}_{0}=\exp (2\pi
i\hat{v}/\hbar )$ and also that $\hat{D}_{0}^{q}=\hat{D}_{2}$. Then, the $q$
states $\Psi _{b,{\bf w},x_{{\rm c}}}^{(d)}=\hat{D}_{0}^{d}\Psi _{b,{\bf w},
x_{{\rm c}}}$, $d=0,...,\ q-1$, are different eigenstates of $\hat{U}
^{\prime }(x_{{\rm c}})$, all associated with the {\em same} eigenvalue.
Each band is therefore $q$-fold degenerate.\newline

The $p$ vectors $\left\{ \exp (imw_{2})\phi _{b}(m;\ {\bf w};\ x_{{\rm c}
})\right\} _{m=0}^{p-1}$, $b=1,...,\ p$, are the eigenvectors of a $p\times p
$ ${\bf w}$-dependent matrix ${\bf M}({\bf w};\ x_{{\rm c}})$,
\begin{equation}
{\bf M}({\bf w};\ x_{{\rm c}})={\bf M}^{(+)}({\bf w};\ x_{{\rm c}}){\bf M}
^{(-)}({\bf w};\ x_{{\rm c}}),  \label{Mw}
\end{equation}
where ${\bf M}^{(\pm )}({\bf w};\ x_{{\rm c}})$ are matrices corresponding
to the operators $\hat{U}_{{\rm KHM}}^{(\pm )}$ in Rel. (\ref{UKHM}) and
whose elements can be explicitly written using the results in Ref. \cite{d2}: 
\begin{equation}
{\bf M}_{m,m^{\prime }}^{(\pm )}({\bf w};\ x_{{\rm c}})=\exp [i\mu \cos (x_{
{\rm c}}\mp w_{1}\mp m^{\prime }\hbar )]\sum_{g=-\infty }^{\infty }J_{\mp
(gp+m^{\prime }-m)}(\mu ,\ x_{{\rm c}})\exp (igpw_{2}),  \label{Mwe}
\end{equation}
where $m,\ m^{\prime }=0,\dots ,\ p-1$ and $J_{m}(\mu ,\ x_{{\rm c}})=(2\pi
)^{-1}\int_{0}^{2\pi }\exp [im(x_{{\rm c}}-u)+i\mu \cos (u)]du$. It is clear
from Eq. (\ref{Mwe}) that the matrix (\ref{Mw}), and therefore the QE
spectrum, is $2\pi $-periodic in $\hbar $ at fixed $\mu $. In addition,
the matrix (\ref{Mw}) is invariant under the ``reflection'' $\hbar 
\rightarrow 2\pi -\hbar $ at fixed $\mu$ provided $x_{{\rm c}}$ and
$w_{1}$ satisfy the two relations 
\begin{equation}
x_{{\rm c}}\mp w_{1}=-x_{{\rm c}}\pm w_{1}+2\pi j_{\pm }\   \label{xcw}
\end{equation}
for some integers $j_{\pm }$. The solution of Eqs. (\ref{xcw}) for
$x_{{\rm c}}$ is $x_{{\rm c}}=\pi (j_{+}+j_{-})/2$. Since $x_{{\rm c}}$ can 
be restricted to $[0,\ \pi /2]$ (see Sec. A), we see that the QE spectrum
is invariant under $\hbar \rightarrow 2\pi -\hbar $ at fixed $\mu $ only
if $x_{{\rm c}}=0,\ \pi /2$.

\begin{center}
{\bf C. General quantum evolution in terms of periodic-wave-packet 
evolutions}\\[0pt]
\end{center}

Following Ref. \cite{kp}, let us express the momentum $v$ in the form 
$v=(n+\beta )\hbar $, where $n$ and $\beta $ are, respectively, the integer
and fractional parts of $v/\hbar $. The $(v,\ x_{{\rm c}})$ representation
of a general 2D wave packet $|\Phi \rangle $ for the system (\ref{H}) will
be then denoted by $\varphi (n+\beta ,\ x_{{\rm c}})$. From the relation
between the $(u,\ x_{{\rm c}})$ and the $(v,\ x_{{\rm c}})$ representations,
one finds that 
\begin{equation}
\Phi (u,\ x_{{\rm c}})=\frac{1}{\hbar }\int \varphi (v/\hbar ,\ x_{{\rm c}
})\exp (iuv/\hbar )dv=\int_{0}^{1}d\beta \exp (i\beta u)\Phi _{\beta }(u,\
x_{{\rm c}}),  \label{fib}
\end{equation}
where $\Phi (u,\ x_{{\rm c}})$ is the $(u,\ x_{{\rm c}})$ representation of 
$|\Phi \rangle $ and 
\begin{equation}
\Phi _{\beta }(u,\ x_{{\rm c}})=\sum_{n=-\infty }^{\infty }\varphi (n+\beta
,\ x_{{\rm c}})\exp (inu).  \label{Pb}
\end{equation}
Clearly, $\Phi _{\beta }(u,\ x_{{\rm c}})$ is $2\pi $-periodic in $u$, so
that Rel. (\ref{fib}) provides the decomposition of $\Phi (u,\ x_{{\rm c}})$
into Bloch functions $\exp (i\beta u)\Phi _{\beta }(u,\ x_{{\rm c}})$ with
quasimomentum $\beta $. Now, since the evolution operator $\hat{U}(\hat{u},\ 
\hat{v};\ x_{{\rm c}})$ in Eq. (\ref{U4}) is $2\pi $-periodic in both 
$\hat{u}$ and $\hat{v}$, its application to such a Bloch function must
``conserve'' $\beta $; in fact, it is easy to see that 
\begin{equation}
\hat{U}(\hat{u},\ \hat{v};\ x_{{\rm c}})[\exp (i\beta u)\Phi _{\beta }(u,\
x_{{\rm c}})]=\exp (i\beta u)\Phi _{\beta }^{\prime }(u,\ x_{{\rm c}}),
\label{cb}
\end{equation}
where 
\begin{equation}
\Phi _{\beta }^{\prime }(u,\ x_{{\rm c}})=\hat{U}\left( \hat{u}=u,\ \hat{v}
=\hbar \beta -i\hbar \frac{d}{du};\ x_{{\rm c}}\right) \Phi _{\beta }(u,\ x_{
{\rm c}}).  \label{Pbp}
\end{equation}
The right-hand side of Eq. (\ref{cb}) is also a Bloch function with
quasimomentum $\beta $, since $\Phi _{\beta }^{\prime }(u,\ x_{{\rm c}})$ is 
$2\pi $-periodic in $u$ due to Eq. (\ref{Pbp}). Rel. (\ref{fib}) then
implies that the evolution of a general wave packet $\Phi (u,\ x_{{\rm c}})$
under $\hat{U}(x_{{\rm c}})$ can be decomposed or ``fibrated'' \cite{kp}
into independent evolutions at fixed $\beta $, 
\begin{equation}
\hat{U}(x_{{\rm c}})\Phi (u,\ x_{{\rm c}})=\int_{0}^{1}d\beta \exp (i\beta u)
\hat{U}_{\beta }(x_{{\rm c}})\Phi _{\beta }(u,\ x_{{\rm c}}),  \label{fibe}
\end{equation}
where $\hat{U}_{\beta }(x_{{\rm c}})$ is the operator appearing on the
right-hand side of Eq. (\ref{Pbp}). Rel. (\ref{fibe}) shows that the quantum
dynamics (\ref{exc}) can be fully reproduced from that of periodic wave
packets (\ref{Pb}) under the corresponding operators $\hat{U}_{\beta }(x_{
{\rm c}})$.\newline

We denote by $\varphi (v/\hbar ,\ x_{{\rm c}};\ t)=\hat{U}^{t}(x_{{\rm c}
})\varphi (v/\hbar ,\ x_{{\rm c}})$ the wave packet at time $t$ in the $(v,\
x_{{\rm c}})$ representation and restrict our attention, as in Sec. IV C, to
the case in which one has a separation of the $(v,\ x_{{\rm c}})$ variables, 
$\varphi (v/\hbar ,\ x_{{\rm c}})=\chi (x_{{\rm c}})\psi (v/\hbar )$, with
normalized $\chi $ and $\psi $. The expectation value $\left\langle \hat{v}
^{2}\right\rangle _{t}$ of $\hat{v}^{2}$ in $\varphi (v/\hbar ,\ x_{{\rm c}
};\ t)$ can then be expressed as follows: 
\begin{equation}
\left\langle \hat{v}^{2}\right\rangle _{t}=\int v^{2}|\psi (v/\hbar ,\ x_{
{\rm c}};\ t)|^{2}dv=\int_{0}^{1}d\beta N_{\beta }\left\langle \hat{v}
^{2}\right\rangle _{\beta ,t},  \label{v2t}
\end{equation}
where $\psi (v/\hbar ,\ x_{{\rm c}};\ t)=\hat{U}^{t}(x_{{\rm c}})\psi
(v/\hbar )$, 
\begin{equation}
\left\langle \hat{v}^{2}\right\rangle _{\beta ,t}=\frac{\hbar ^{2}}{N_{\beta
}}\sum_{n=-\infty }^{\infty }(n+\beta )^{2}|\psi (n+\beta ,\ x_{{\rm c}};\
t)|^{2},\ \ \ \ \ \ N_{\beta }=\sum_{n=-\infty }^{\infty }|\psi (n+\beta
)|^{2},  \label{v2bt}
\end{equation}
and $\psi (n+\beta ,\ x_{{\rm c}};\ t)=$ $\hat{U}_{\beta }^{t}(x_{{\rm c}
})\psi (n+\beta )$. We notice that the expressions (\ref{v2t}) and 
(\ref{v2bt}) are independent of $\chi (x_{{\rm c}})$ and their 
$x_{{\rm c}}$-dependence is completely due to that of $\hat{U}(x_{{\rm c}})$ 
or $\hat{U}_{\beta }(x_{{\rm c}})$. Fully analogous expressions can be 
written for the expectation values $\left\langle \hat{u}^{2}\right\rangle 
_{t}$ and $\left\langle \hat{u}^{2}\right\rangle _{\gamma ,t}$, where 
$\gamma $ is the quasimomentum characterizing Bloch functions $2\pi
$-quasiperiodic in the $v$ direction. Let us now assume approximate
quantum diffusions of $\left\langle \hat{v}^{2}\right\rangle _{\beta ,t}$ and 
$\left\langle \hat{u} ^{2}\right\rangle _{\gamma ,t}$ for all $(\beta ,\
x_{{\rm c}})$, $(\gamma ,\ x_{{\rm c}})$, and for sufficiently large $t$,
i.e., $\left\langle \hat{v} ^{2}\right\rangle _{\beta ,t}\approx
2D_{\beta }(x_{{\rm c}})t$ and $\left\langle \hat{u}^{2}\right\rangle 
_{\gamma ,t}\approx 2D_{u,\gamma }(x_{{\rm c}})t$. The expectation value of 
the kinetic energy then exhibits the approximate diffusive behavior
$\left\langle \omega ^{2}(\hat{u}^{2}+\hat{v} ^{2})/2\right\rangle
_{t}\approx 2D_{{\rm KE}}(x_{{\rm c}})t$, where 
\[
D_{{\rm KE}}(x_{{\rm c}})=(\omega ^{2}/2)\left[ \int_{0}^{1}d\beta N_{\beta
}D_{\beta }(x_{{\rm c}})+\int_{0}^{1}d\gamma N_{u,\gamma }D_{u,\gamma }(x_{
{\rm c}})\right] .
\]
Following the same reasoning as in the case of $\beta =0$ considered in Sec.
IV C, the diffusion coefficients $D_{\beta }(x_{{\rm c}})$, $D_{u,\gamma
}(x_{{\rm c}})$, and $D_{{\rm KE}}(x_{{\rm c}})$ for sufficiently small $K$
are expected to satisfy approximately formula (\ref{Dxc}) with $D_{{\rm H}
}^{(\pm )}$ replaced by corresponding proportionality constants 
$D_{{\rm H},\beta }^{(\pm )}$, $D_{{\rm H},u,\gamma }^{(\pm )}$, and 
\[
D_{{\rm H},{\rm KE}}^{(\pm )}=(\omega ^{2}/2)\left[ \int_{0}^{1}d\beta
N_{\beta }D_{{\rm H},\beta }^{(\pm )}+\int_{0}^{1}d\gamma N_{u,\gamma }
D_{{\rm H},u,\gamma }^{(\pm )}\right] .
\]

\begin{center}
{\bf D. Cases in which }$D_{{\rm H},\beta }^{(+)}=|D_{{\rm H},\beta }^{(-)}|
${\bf \\[0pt]}
\end{center}

We show here that there are cases in which $D_{{\rm H},\beta }^{(+)}=|D_{
{\rm H},\beta }^{(-)}|$ (similarly, one can show that there are cases in
which $D_{{\rm H},u,\gamma }^{(+)}=|D_{{\rm H},u,\gamma }^{(-)}|$). The
constants $D_{{\rm H},\beta }^{(+)}$ and $|D_{{\rm H},\beta }^{(-)}|$ are
the diffusion coefficients for $\left\langle \hat{v}^{2}\right\rangle
_{\beta ,t}$ under the standard Harper evolution operator $\hat{U}_{{\rm H}
}=\exp \left\{ i\left[ \cos (\hat{u})+\cos (\hat{v})\right] /\hbar \right\} $
and its inverse $\hat{U}_{{\rm H}}^{-1}$, respectively. In the $u$
representation, 
\begin{equation}
\hat{U}_{{\rm H}}^{t}(u)=\exp \left\{ \frac{it}{\hbar }\left[ \cos (u)+\cos
\left( i\hbar \frac{d}{du}\right) \right] \right\} .  \label{UtH}
\end{equation}
Consider the Bloch state $\Lambda _{\beta }(u)=\exp (i\beta u)\Phi (u)$,
where $\Phi (u)$ is given by the right-hand side of Eq. (\ref{pcs}), and
define the variable $u^{\prime }=2\overline{u}-u$. It is easy to see that 
\begin{equation}
\Lambda _{\beta }(u)=C\Lambda _{\beta }^{\ast }(u^{\prime }),  \label{cc}
\end{equation}
where $C=\exp [2i\overline{u}(\overline{v}+\beta )/\hbar ]$ is a constant
phase factor. Choosing now $\overline{u}=j\pi $, $j$ integer, and using
Rels. (\ref{UtH}) and (\ref{cc}), we find that 
\begin{equation}
\Lambda _{\beta }(u;\ -t)\equiv \hat{U}_{{\rm H}}^{-t}(u)\Lambda _{\beta
}(u)=C\left[ \hat{U}_{{\rm H}}^{t}(u^{\prime })\Lambda _{\beta }(u^{\prime })
\right] ^{\ast }=C\Lambda _{\beta }^{\ast }(u^{\prime };\ t).  \label{tmt}
\end{equation}
It follows from Rel. (\ref{tmt}) that 
\begin{equation}
\left\langle \hat{v}^{2}\right\rangle _{\beta ,-t}=-\hbar ^{2}\int_{0}^{2\pi
}\Lambda _{\beta }^{\ast }(u;\ -t)\frac{d^{2}}{du^{2}}\Lambda _{\beta }(u;\
-t)du=\left\langle \hat{v}^{2}\right\rangle _{\beta ,t}^{\ast }=\left\langle 
\hat{v}^{2}\right\rangle _{\beta ,t},  \label{vbtmt}
\end{equation}
since $\left\langle \hat{v}^{2}\right\rangle _{\beta ,t}$ must be real (and
positive). The equality $D_{{\rm H},\beta }^{(+)}=|D_{{\rm H},\beta }^{(-)}|$
is an immediate consequence of Rel. (\ref{vbtmt}).\newline

\newpage Table I. $D(0)/K$ and $D(\pi /2)/K^{2}$ for several values of $K$. 
\newline
\[
\begin{tabular}{|l|l|l|l|l|l|l|}
\hline
$K$ & $0.15$ & $0.157$ & $0.17$ & $0.18$ & $0.19$ & $0.2$ \\ \hline
$D(0)/K$ & $0.458$ & $0.459$ & $0.461$ & $0.461$ & $0.460$ & $0.459$ \\ 
\hline
$D(\pi /2)/K^{2}$ & $0.219$ & $0.226$ & $0.237$ & $0.243$ & $0.244$ & $0.243$
\\ \hline
\end{tabular}
\]

\figure{FIG. 1. Classical stochastic webs for $K=0.157$ and $\alpha
=\pi/2$, generated by the map (\ref{M}) in three cases: (a) $x_{{\rm
c}}=0$; (b) $x_{{\rm c}}=1.47$; (c) $x_{{\rm c}}=\pi /2$. The range of
both $u$ and $v$ in all the plots is $[-2\pi ,\ 2\pi ]$.\label{f1}}

\figure{FIG. 2. ``Hofstadter butterflies" (HBs) $E(\hbar )$ for $\mu
=0.5$ and: (a) $x_{{\rm c}}=0$, $-4\leq E\leq 4$; (b) $x_{{\rm c}}=1.47$,
$-1\leq E\leq 1$; (c) $x_{{\rm c}}=\pi /2$, $-1\leq E\leq 1$.\label{f2}}

\figure{FIG. 3. Perturbed double HB $E'(\hbar )$ ($-2\leq E' \leq 2$),
obtained by locally scaling the HB in Fig. 2(c) by $\epsilon
^{-1}=[\mu\sin (\hbar /2)]^{-1}$. The perturbation is reflected by the
fact that the spectral width $\Delta E'(\pi )$ (at the HB center) is
slightly smaller than the maximal width $\Delta E'(2\pi )=4$.\label{f3}}

\figure{FIG. 4. Solid lines: $\left\langle \hat{v}^{2}\right\rangle _{t}$
for $\hbar /(2\pi )=[51+(\sqrt{5}-1)/2]^{-1}$, $K=0.157$, and $x_{{\rm
c}}=0$ (upper line), $x_{{\rm c}}=1.47$ (middle line), $x_{{\rm c}}=\pi
/2$ (lower line). Dashed lines: Corresponding classical quantity
$\left\langle v^{2}_{t}\right\rangle$ for the same values of $x_{{\rm
c}}$ (see text for more details). For the sake of visibility, the lines
for $x_{{\rm c}}=\pi /2$ were shifted below by multiplying both
$\left\langle \hat{v}^{2}\right\rangle _{t}$ and $\left\langle
v^{2}_{t}\right\rangle$ by $0.01975$. The inset shows the normal plot of
$\left\langle \hat{v}^{2}\right\rangle _{t}$ for $x_{{\rm c}}=\pi /2$.
The values of $\hbar$ and $K$ above were used also to obtain the results
in Figs. 5 and 6.\label{f4}}

\figure{FIG. 5. Diamonds: Numerical results for the quantum-diffusion
coefficient $D(x_{{\rm c}})$ (see text for more details). Solid line:
Least-square fit of formula (\ref{Dxc}) to the numerical data, $D_{{\rm
H}}^{(+)}=0.224$.\label{f5}}

\figure{FIG. 6. Middle line: Global quantum diffusion of
$\overline{\left\langle \hat{v}^{2}\right\rangle _{t}}$, given by a
uniform average of $\left\langle \hat{v}^{2}\right\rangle _{t}$ over
$x_{{\rm c}}=k\pi /40$, $k=0,\ 1,\dots ,\ 18$ (see text for more
details). The upper and lower lines correspond to $\left\langle
\hat{v}^{2}\right\rangle _{t}$ for $x_{{\rm c}}=0$ ($k=0$) and $x_{{\rm
c}}=2\pi /5$ ($k=16$), respectively.\label{f6}}


\begin{references}
\bibitem{note} The spectral width $\Delta E^{\prime }(\pi )$ of
$\hat{H}_{{\rm eff}}/\epsilon $ can calculated exactly by diagonalizing
a simple $2\times 2$ matrix (\ref{Mw}) with (\ref{Mwe}), while $\Delta
E^{\prime }(2\pi )$ is the width of the single band of $\hat{H}_{{\rm
eff}}/\epsilon = \hat{H}_{1}+\epsilon \hat{H}_{2}+\dots $ for $\hbar
=2\pi $. Since $\epsilon =0$ for $\hbar =2\pi $, $\hat{H}_{{\rm eff}}/
\epsilon $ is exactly equal to $\hat{H}_{1}$, the Harper Hamiltonian
(\ref{H1s}) for which $\Delta E^{\prime }(2\pi )=4$, independent of $\mu $.
\end{references}
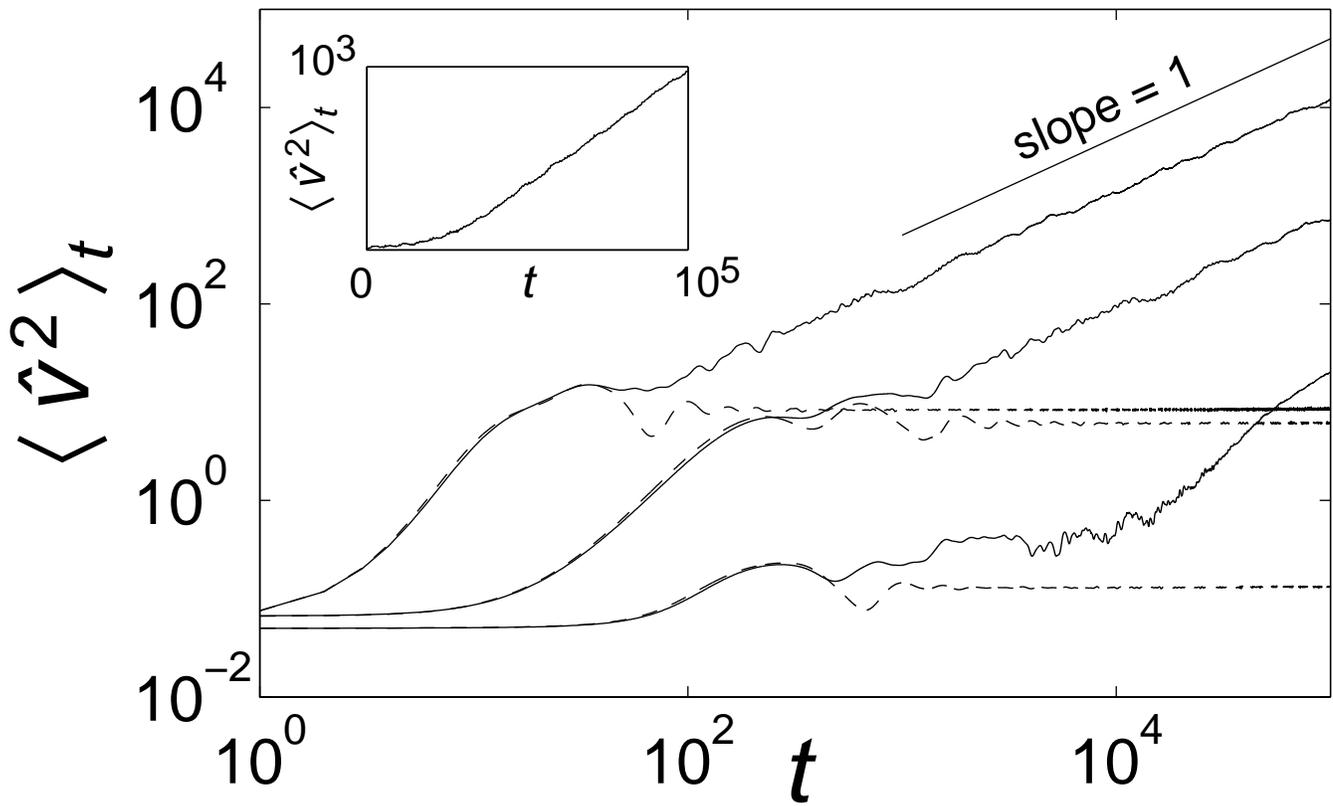
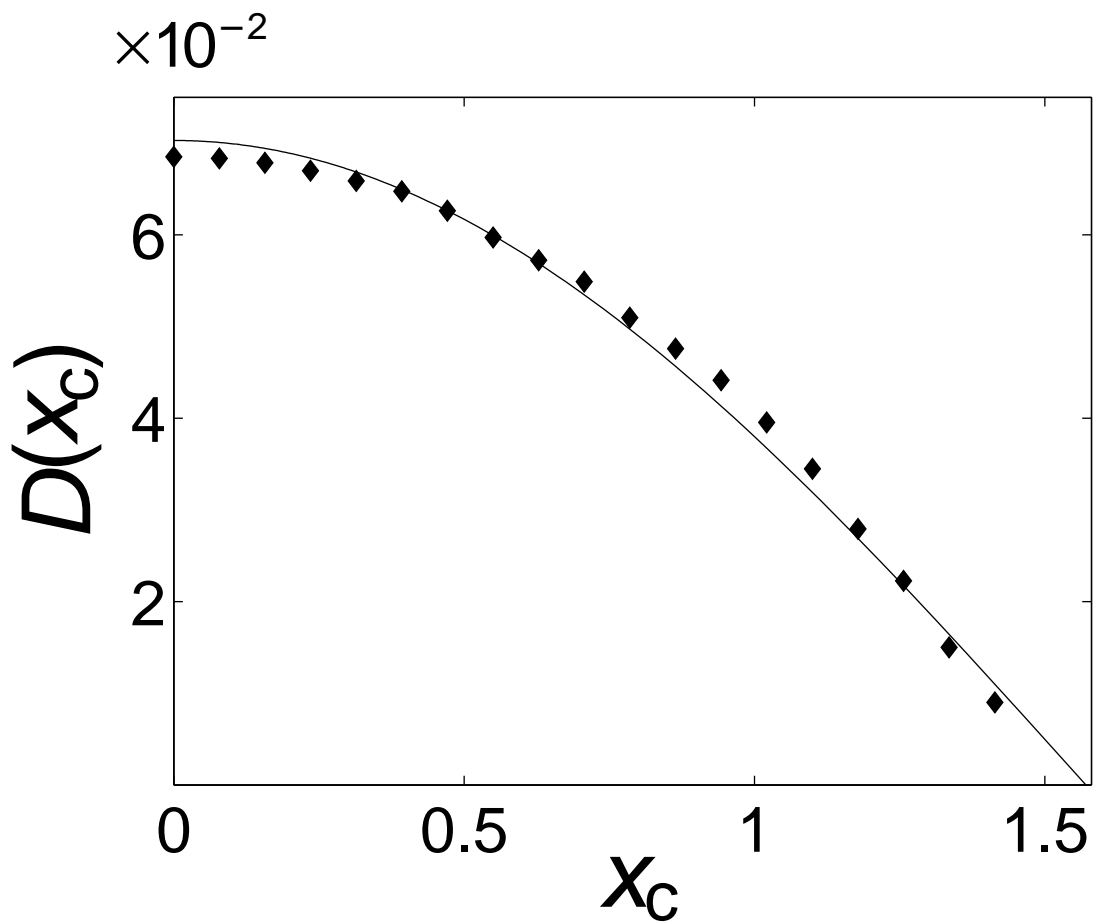
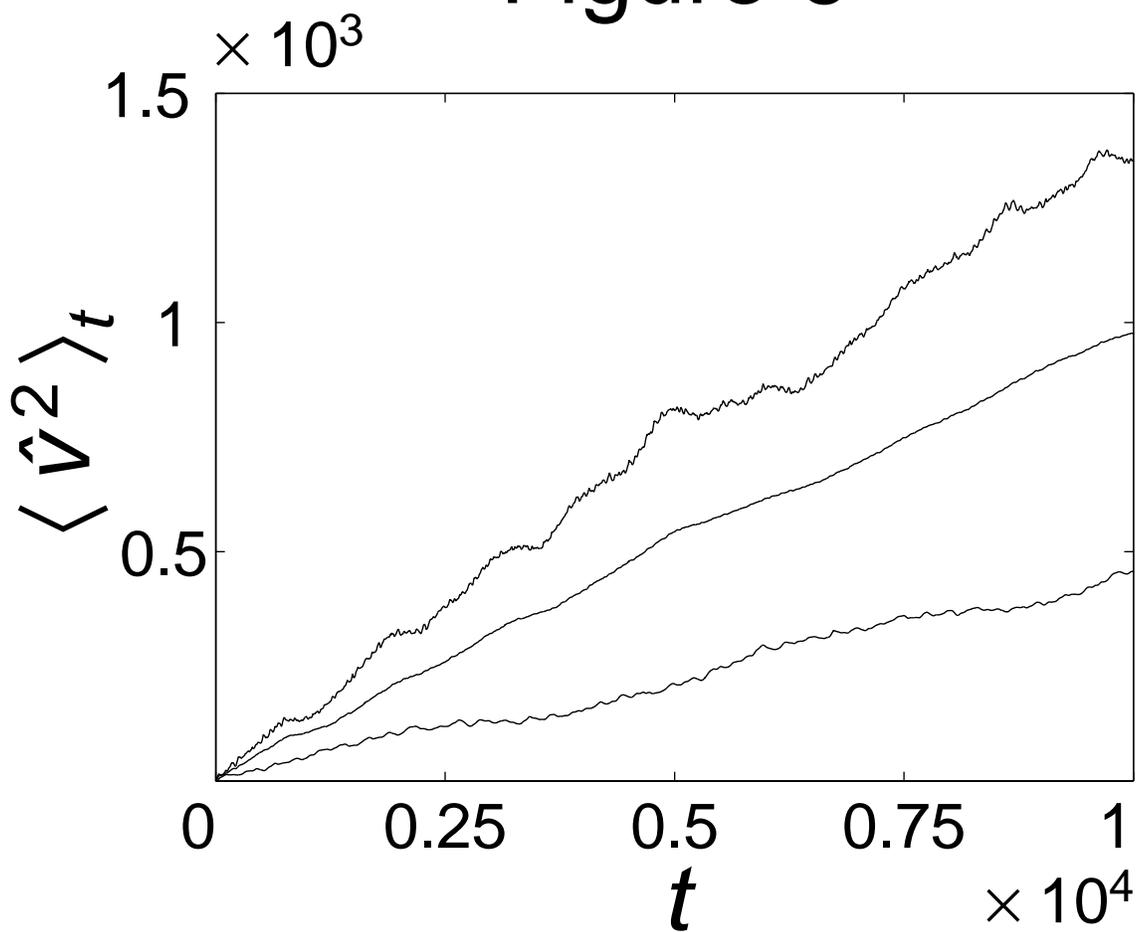
\end{document}